\documentstyle[own_sngl]{article}
\newcommand{\oversim}[2]{\protect{\mbox{\lower0.5ex\vbox{%
   \baselineskip=0pt\lineskip=0.2ex
   \ialign{$\mathsurround=0pt #1\hfil##\hfil$\crcr#2\crcr\sim\crcr}}}}} 
\newcommand{\simgreat}{\mbox{$\,\mathrel{\mathpalette\oversim>}\,$}} 
\newcommand{\simless} {\mbox{$\,\mathrel{\mathpalette\oversim<}\,$}} 
\slugcomment{{\em {\bf NewA}, in press, 27.07.99}}
\lefthead{Kroupa}
\righthead{ONC Solutions}
\begin{document}
%
\title {Constraints on Stellar-Dynamical Models of the Orion Nebula
Cluster}

\author {Pavel Kroupa\\
\medskip
\small{MS42, Harvard-Smithsonian Centre for Astrophysics, 60~Garden
Street, Cambridge, MA~02138, USA\\ e-mail: pkroupa@cfa.harvard.edu \\
(On leave from the Institut f{\"u}r Theoretische Astrophysik,
Universit{\"a}t Heidelberg) }}

\begin{abstract}
\noindent 
The results obtained by Kroupa, Petr \& McCaughrean (1999) for
specific models of young compact binary-rich clusters are generalised
using dynamical scaling relations, to infer the candidate set of
possible birth models leading to the Orion Nebula Cluster (ONC), of
which the Trapezium Cluster is the core.  It is found that candidate
sets of solutions exist which allow the ONC to be in virial
equilibrium, expanding or contracting. The range of possible solutions
is quite narrow.

These results will serve as guidelines for future, CPU-intensive
calculations of the stellar-dynamical and astrophysical evolution of
the entire ONC. These, in turn, will be essential to quantify
observables that will ultimately discriminate between models, thus
allowing us to understand if the ONC is in the process of assembling a
rich Galactic cluster, and, if this is the case, how it occurs.
\end{abstract}


\section{INTRODUCTION} 
\label{sec:intro}
\noindent
The formation of bound star clusters is a major unsolved problem in
observational and theoretical astrophysics. Significant observational
(e.g. Lada \& Lada 1991; Megeath et al. 1996; Lada, Alves \& Lada
1996; Lada, Falgarone \& Evans 1997) and theoretical (e.g. Lada,
Margulis \& Dearborn 1984; Elmegreen \& Efremov 1997; Klessen, Burkert
\& Bate 1998) efforts are being directed at solving this problem.
There are serious technical difficulties. For example, on the
observational side, the obscuration of a part of the very young
stellar and proto-stellar population and significant uncertainties in
pre-main sequence evolution (e.g. Wuchterl \& Tscharnuter 1999) hamper
the understanding of the spatial and age distribution within compact
star-forming regions.  On the theoretical side, the presently next to
impossible handling of three-dimensional self-gravitating
magneto-hydrodynamics with feedback from forming stars limits detailed
understanding of the onset of star-formation in general, and cluster
formation in particular.

This paper forms part of a theoretical approach based on purely
stellar-dynamical arguments. These are limited to the phase after the
proto-stars and ambient gas decouple dynamically, but allow insights
into the evolution of a stellar assemblage after the massive stars
rapidly remove the remaining gas. Understanding this phase of the
evolution can provide a lead towards the initial conditions that give
rise to bound Galactic clusters.

Stellar-dynamical calculations of binary-rich young compact clusters
with Aarseth's {\sc Nbody5} code were presented by Kroupa, Petr \&
McCaughrean (1999, hereinafter KPM), to study the evolution of the
binary population and of the cluster bulk properties.  The results
were applied to a comparison with key observables available for the
Trapezium Cluster, and a solution was found in which the Trapezium
Cluster, with a stellar mass of $700\,M_\odot$, is expanding rapidly
due to the expulsion of about $1000\,M_\odot$ in gas roughly
$50\,000$~yr ago. However, while that work quantifies the evolution of
the binary population and of the cluster within a few~Myr, the
question remains open as to how unique the expanding solution of the
Trapezium Cluster is.

In this paper the solution space of all possible models of the ONC is
studied by invoking stellar-dynamical scaling relations that capture
the gross behaviour of star clusters, and by building on the results
obtained by KPM.  Section~\ref{sec:obs} summarises the key observables
used to constrain the allowed range of models, and in
Section~\ref{sec:gen} the KPM results are generalised and the
candidate solution space is identified. The conclusions follow in
Section~\ref{sec:conc}.

\section{OBSERVATIONAL CONSTRAINTS}
\label{sec:obs}
\noindent 
A detailed description of the observational constraints can be found
in KPM. Here a brief summary is given.  The key observables used to
constrain the stellar-dynamical models by KPM are the central stellar
number density, the pre-main-sequence age of the stellar population,
the velocity dispersion, and the binary proportion.

Based on a high-spatial-resolution direct-imaging near-infrared study,
McCaughrean \& Stauffer (1994) calculate that 29~systems are within
the central spherical volume with a radius of $R=0.053$~pc. This
corresponds to a central number density $\rho_{\rm c}^{\rm (obs)} =
4.65\times10^4$~stars/pc$^3$ (strictly speaking systems/pc$^3$, but
the maximal factor of~2 error is not critical in this analysis).  In
a near-infrared study covering a field of $\sim 0.65\,{\rm pc} \times
0.65\,{\rm pc}$ centred on the Trapezium Cluster, McCaughrean et
al. (1996) count 700 systems.  

From the analysis of the HR-Diagram, based on optical photometry and
spectroscopy, Prosser et al. (1994), Hillenbrand (1997) and Palla \&
Stahler (1999) derive a mean age for the whole Orion Nebula Cluster of
less than about 1~Myr.

The one-dimensional velocity dispersion within~0.41~pc of the centre
of the Trapezium Cluster is derived by Jones \& Walker (1988), from
their relative proper-motion survey using photographic plates, to be
$\sigma^{\rm (obs)}=2.54\pm0.27$~km/s. Because the plate-reduction
algorithms eliminate any signature due to rotation and/or expansion or
contraction, it is unknown from that work and others (see KPM) if the
Trapezium Cluster is expanding or contracting. Frink, Kroupa \&
R\"oser (1999) measure the bulk motions in the ONC using new absolute
proper motion data, and find some expanding as well as contracting
motion. Their result is consistent with the ONC currently expanding in
an approximately cylindrical molecular cloud potential (Kroupa \&
Frink 1999).  Following KPM the velocity dispersion obtained from the
more precise relative proper motions is used in the present analysis.

High-resolution imaging using a variety of techniques has shown that
the binary proportion within about 0.3~pc of the centre of the
Trapezium Cluster is significantly below that seen in Taurus--Auriga,
but consistent with the Galactic field value (Prosser et al. 1994;
Petr 1998; Petr et al. 1998). This may be the result of disruption
through encounters in the cluster (KPM).

The Trapezium Cluster contains virtually no gas (Wilson et al. 1997),
and a large part of the ONC population is optically visible
(Hillenbrand \& Hartmann 1998), so that purely stellar-dynamical
calculations can be applied. 

There is little evidence for sub-structure in the ONC (Bate, Clarke \&
McCaughrean 1998). 

\section{THE KPM STELLAR-DYNAMICAL MODELS}
\label{sec:mods}
\noindent
Details are found in KPM.  The KPM models contain $N=1600$~stars
(cluster mass $M_{\rm cl}=700\,M_\odot$), with a Galactic-field IMF
with stellar masses in the range $0.08\,M_\odot$ to $30\,M_\odot$. The
stars are paired at random to form a primordial binary proportion of
unity, $f_{\rm tot}=1$ with a period distribution as seen in
Taurus--Auriga, and $f_{\rm tot}=0.6$ with a period distribution as in
the Galactic field. Here, the total (counting all separations) binary
proportion is $f_{\rm tot}=N_{\rm bin}/(N_{\rm bin}+N_{\rm sing})$,
where $N_{\rm bin}$ and $N_{\rm sing}$ are the number of binary and
single-star systems, respectively.  Stellar evolution is not treated,
and the dynamical evolution is followed for 5~Myr only.

Six different models are constructed to cover a range of possible
cases, each having initially a Plummer density profile with 
half-mass radius $R_{0.5}$:
\begin{itemize}
\item Virial equilibrium models with primordial $f_{\rm tot}=1$
(model~A1) and $f_{\rm tot}=0.6$ (model~A2). These have initially 
$R_{0.5}=0.1$~pc, $t_{\rm cross}=0.06$~Myr (eqn.~\ref{eqn:tcr}) and 
$t_{\rm relax}=0.62$~Myr (eqn.~\ref{eqn:tr}). 
\itemsep -1mm
\item Expanding clusters assuming a star formation efficiency
$\epsilon=M_{\rm cl}/(M_{\rm cl}+M_{\rm gas})=0.42$ and instantaneous
loss of a gas mass $M_{\rm gas}=967\,M_\odot$ at the start of
the computation, with binary proportions $f_{\rm tot} =1$ (model~B1)
and~0.6 (model~B2). These models have $R_{0.5}=0.1$~pc and an initial
velocity dispersion that is $\epsilon^{-1/2}$ larger than the virial
equilibrium value without the gas.  
\itemsep -1mm
\item Clusters that collapse with an initial velocity
dispersion that is a factor~0.14 times smaller than the virial
equilibrium value, with initially $R_{0.5}=0.4$~pc (model~C1)
and~0.8~pc (model~C2). These have $f_{\rm tot}=1$. 
\end{itemize}
Three renditions of each model yield averages of the relevant observables.

The choice for $N$ is based on the observed number of systems in the
Trapezium Cluster. The initial central density ($\rho_{\rm c,
in}=10^{5.92}$ stars/pc$^3$) of the virial equilibrium and expanding
models is larger than the presently observed value ($\rho_{\rm c}^{\rm
(obs)}=10^{4.67}$ stars/pc$^3$), because a reduction of the central
density with time is anticipated without knowing the detailed
evolution in advance, since no such compact and binary-rich clusters
had ever been evolved until then. For the collapsing models, the
initial possible configuration is unknown, since $N$-body collapse
calculations of an initially centrally concentrated density profile
and with a large proportion of realistic binary systems had never been
done before. The values $R_{0.5}=0.4$ and~0.8~pc are thus first tries.

The mean stellar mass is $0.44\,M_\odot$. It is used throughout this
paper: $M_{\rm cl}=0.44\times N\,M_\odot$.

\section{GENERALISATION}
\label{sec:gen}
\noindent
Before constraining the solution space using dynamical arguments, a
pre-selection can be made immediately, by noting that the lower limit
on $N$ is 700~stars, since this many systems have been counted in the
Trapezium Cluster (Section~\ref{sec:obs}).

It is known that the binary fraction is high (about 50--60~per cent),
so $N$ is likely to be closer to 1200. However, since some loss of
stars is expected through dynamical ejection, and in general location
at larger radii than the survey limit, $N=1600$ is deemed more
appropriate to be a lower limit. There is thus little room for an
initial $N$ below about~1600.

Assuming the Trapezium Cluster is merely the core of the ONC, an upper
limit $N=5600-10000$ is obtained for the whole ONC.  The lower value is
valid if the overall binary proportion is $f_{\rm tot}=0.6$, since
about~3500 systems have been counted within approximately 2.5~pc of
the cluster's centre (Hillenbrand \& Hartmann 1998). The upper limit
is estimated by assuming $f_{\rm tot}=1$ as a probably more realistic
value for most of the ONC, implying that 7000~stars will have been
counted, and that 30~per cent of all systems have been lost through
expansion beyond the 2.5~pc radius, or have not been seen because they
are too deeply embedded or too faint.

\subsection{Clusters in virial equilibrium}
\label{sec:gen_vireq}
\noindent
Without doing additional time-consuming numerical experiments, the
evolution of model clusters in virial equilibrium with different $N$
and $R_{0.5}$ can be assessed through simple scaling arguments.

\begin{figure}
\plotfiddle{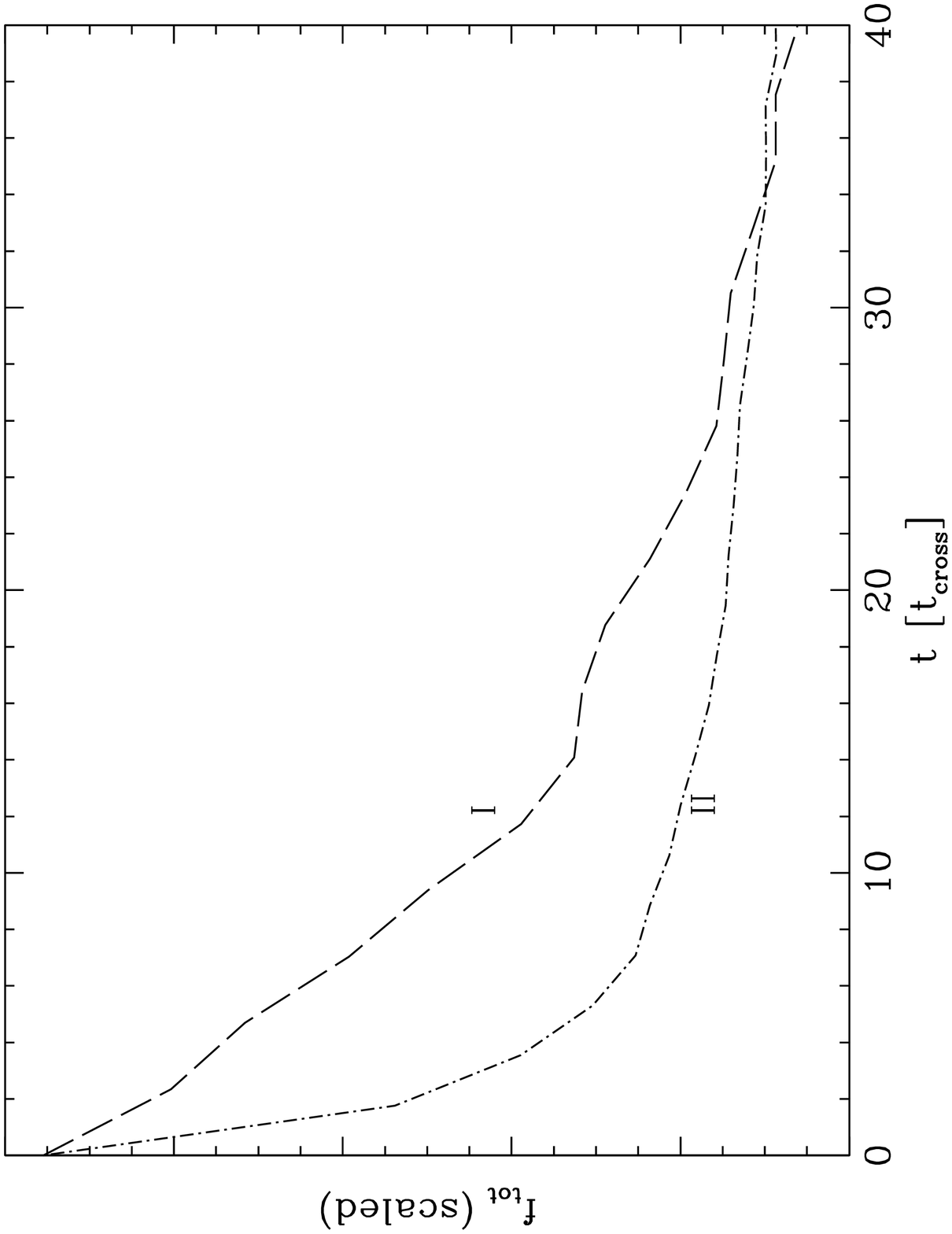}{17cm}{-90}{60}{60}{-250}{370}
\caption{Depletion of the binary proportion in two very different
clusters. The time-scale is in units of the initial crossing time,
whereas the vertical axis is scaled to the minimum and maximum $f_{\rm
tot}$.  The long-dashed curve (I) is the average of five $N$-body
renditions of a cluster with initially 200~binaries and a Plummer
density distribution with $R_{0.5}=2.5$~pc ($\rho_{\rm
c}=13$~stars/pc$^3$), and stars with masses in the range 0.1~to
$1.1\,M_\odot$ (from Kroupa 1995). The initial crossing time is
18~Myr, the initial $f_{\rm tot}=1.00$, and the final $f_{\rm
tot}=0.83$.  The short-dash-dotted curve (II) is for model~A1 in KPM
(and this paper; $\rho_{\rm c}=10^{5.92}$~stars/pc$^3$), for which the
initial $f_{\rm tot}=0.76$ (reduced from unity because of disruption
through crowding, see KPM), the final $f_{\rm tot}=0.34$, and initial
$t_{\rm cross}=0.059$~Myr. For the virial--equilibrium models
discussed in this paper, $f_{\rm tot}(t)$ will lie between the two
extreme cases shown here.
\label{fig:ft_tcr}}
\end{figure}

Binary depletion occurs approximately on a crossing time-scale,
$t_{\rm cross}$ (fig.~3 in KPM), whereas the velocity dispersion and
central density change on a time-scale comparable to the relaxation
time, $t_{\rm relax}$ (fig.~2 in KPM). That complete destruction of
long-period or soft binaries occurs within a few tens of initial
$t_{\rm cross}$ is demonstrated in Fig.~\ref{fig:ft_tcr}, in which the
evolution of $f_{\rm tot}$ for two very different clusters is
compared. Despite very different initial crossing times and different
$N$, $f_{\rm tot}$ decays on a comparable time-scale~$\approx\,{\rm
few}\times t_{\rm cross}$. That the decay is not exactly in-phase does
not affect the conclusions of this paper.

The three-dimensional velocity dispersion, 
\begin{equation}
\sigma = s\,\left({G\,M_{\rm cl}\over 2\,R_{0.5}}\right)^{1\over2}, 
\label{eqn:1dvd}
\end{equation}
where $G=4.49\times10^{-3}\,{\rm pc}^3/(M_\odot\,{\rm Myr}^2)$ is the
gravitational constant, and $s$ is a structure constant of order unity
defined by the details of the density distribution.  The
one-dimensional velocity dispersion is $\sigma_{\rm 1D} =
\sigma/\sqrt{3}$, assuming the velocity distribution is isotropic.
For $M_{\rm cl}=700\,M_\odot$ and $R_{0.5}=0.1$~pc, $\sigma_{\rm
1D}/s=2.29$~pc/Myr (1~pc/Myr$=0.982$~km/s, but throughout this paper
no distinction is made between these two units), and from fig.~2 in
KPM, $\sigma_{\rm 1D}(t=0)=1.95$~km/s, so that $s=0.85$ throughout this
paper.

The typical crossing time, 
\begin{equation}
t_{\rm cross} = {2\,R_{\rm 0.5}\over \sigma},
\label{eqn:tcr}
\end{equation}
whereas the relaxation time in Myr (from Binney \& Tremaine 1987),
\begin{equation}
t_{\rm relax}\,=\,{21\over \ln(0.4\,N)}
   \left({M_{\rm cl} \over 100\,M_\odot}\right)^{1/2} 
   \left({1\,M_\odot\over \overline{m}}\right) 
   \left({R_{0.5} \over 1\,{\rm pc}}\right)^{3/2}~.
\label{eqn:tr}
\end{equation}

Thus, it is immediately apparent that increasing $N$ (i.e. $M_{\rm
cl}$) will, for the same $R_{0.5}$, {\it shorten} the time-scale for
binary depletion while simultaneously {\it slowing down} the
time-scale for the evolution of the central density and velocity
dispersion. This implies that by increasing $N$, the variation of the
central density becomes increasingly negligible over the age of the
cluster. This is already evident in fig.~2 in KPM, where the central
density decays by a factor of~2.5 only during the first~Myr.  The
central density for a Plummer density distribution is
\begin{equation}
\rho_{\rm c} = {3\,N \over 4\,\pi\,\left(0.77\,R_{0.5}\right)^3},
\label{eqn:rhoc}
\end{equation}
where $R_{\rm pl}=0.77\,R_{0.5}$ is the Plummer radius. 

\begin{figure}
\plotfiddle{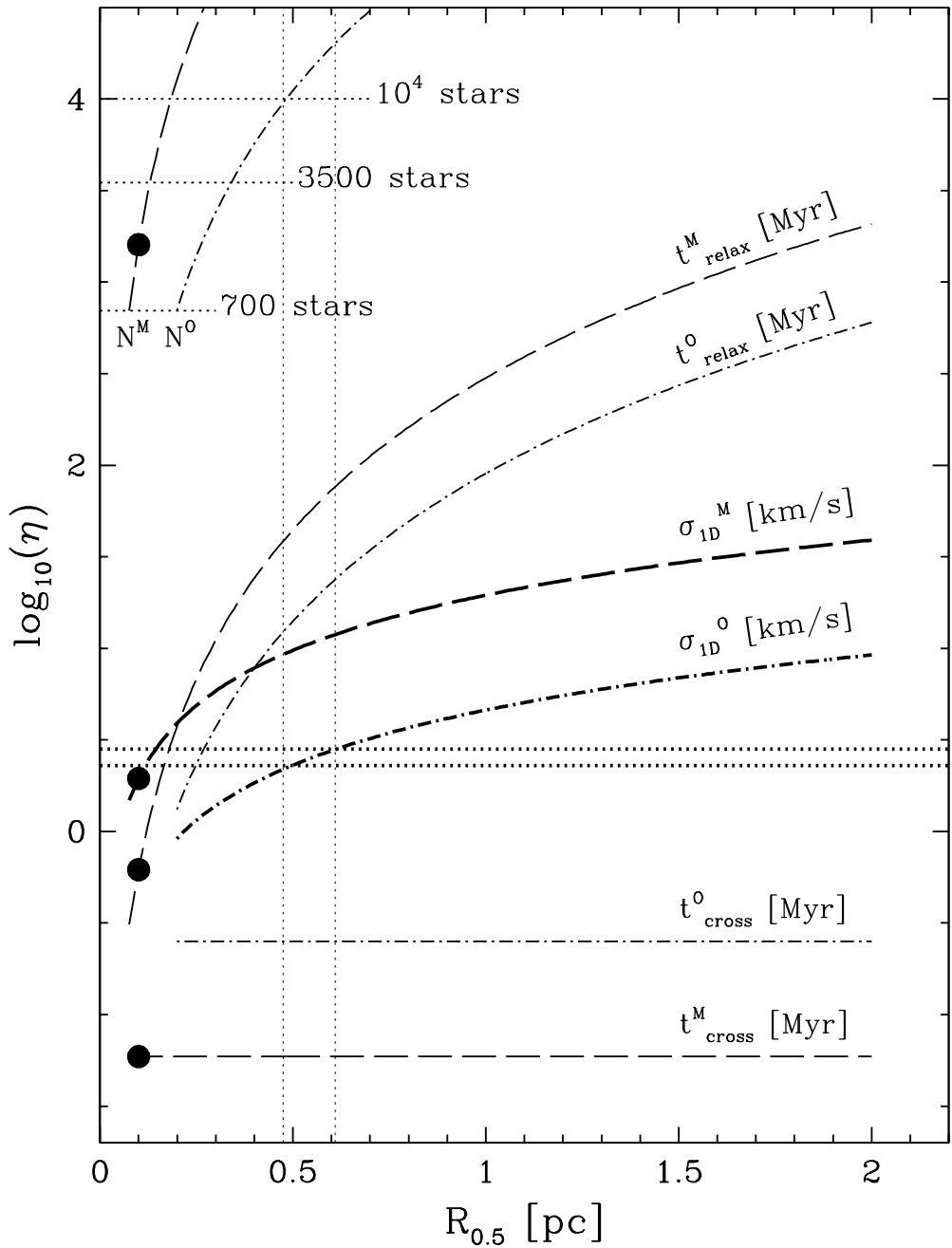}{17cm}{0}{70}{70}{-225}{-15}
\caption{Two sets of initial model clusters in virial equilibrium
indicated by the super-scripts~M (long-dashed curves) and~O
(short-dash-dotted curves). M~refers to \underbar{m}odels that have
the same initial density as models~A, and O~refers to models with a
central density equal to the \underbar{o}bserved value in the
Trapezium Cluster. The number of stars, $\eta=N$, the relaxation time,
$\eta=t_{\rm relax}$, and crossing time, $\eta=t_{\rm cross}$, and the
one-dimensional velocity dispersion, $\eta=\sigma_{\rm 1D}$, are
plotted as log$_{10}[\eta(R_{0.5})]$ for each model. For example,
models~A have initially $R_{0.5}=0.1$~pc with $N=1600$,
log$_{10}(t_{\rm cross})=-1.23$ (Myr), log$_{10}(t_{\rm relax})=-0.21$
(Myr) and log$_{10}(\sigma_{\rm 1D})=0.29$ (km/s). These are
identified by the large black dots. The thin vertical dotted lines
delineate the possible set of equilibrium solutions: O-models with
$0.48<R_{0.5}<0.61$~pc have the same central density as the Trapezium
Cluster, and a velocity dispersion consistent within the one-sigma
uncertainty range with the observed value, which is shown by the two
horizontal dotted lines.
\label{fig:sc_vir}}
\end{figure}

For illustrative purposes, two sets of initial models are constructed
and compared in Fig.~\ref{fig:sc_vir}. One set of models has ${\rm
log}_{10}(\rho_{\rm c})=5.92$ (stars/pc$^3$), which is the same as in
models~A. The other set has ${\rm log}_{10}(\rho_{\rm c})=4.67$
(stars/pc$^3$), which is the same as the observed central density in
the Trapezium Cluster. The first set is clearly not a viable solution
set since the initial density remains too high ($t_{\rm relax}$ is too
long).

However, there is a set of virial-equilibrium candidate solutions if
$10000\simless N\simless 16000$ and $0.48\simless R_{0.5} \simless
0.61$, for the constraint $\rho_{\rm c}=\,$constant.  These models
have a velocity dispersion consistent with the observed value, and
$t_{\rm cross}=2.5\times10^5$~yr, which is sufficiently short for
significant binary depletion to have occurred in the ONC until now
(Fig.~\ref{fig:ft_tcr}).  Fully self-consistent $N$-body calculations
are needed to verify if a Taurus--Auriga-like binary population can be
reduced to the observed level in the Trapezium Cluster. The
calculations of KPM suggest that this is likely to be the case.

\begin{figure}
\plotfiddle{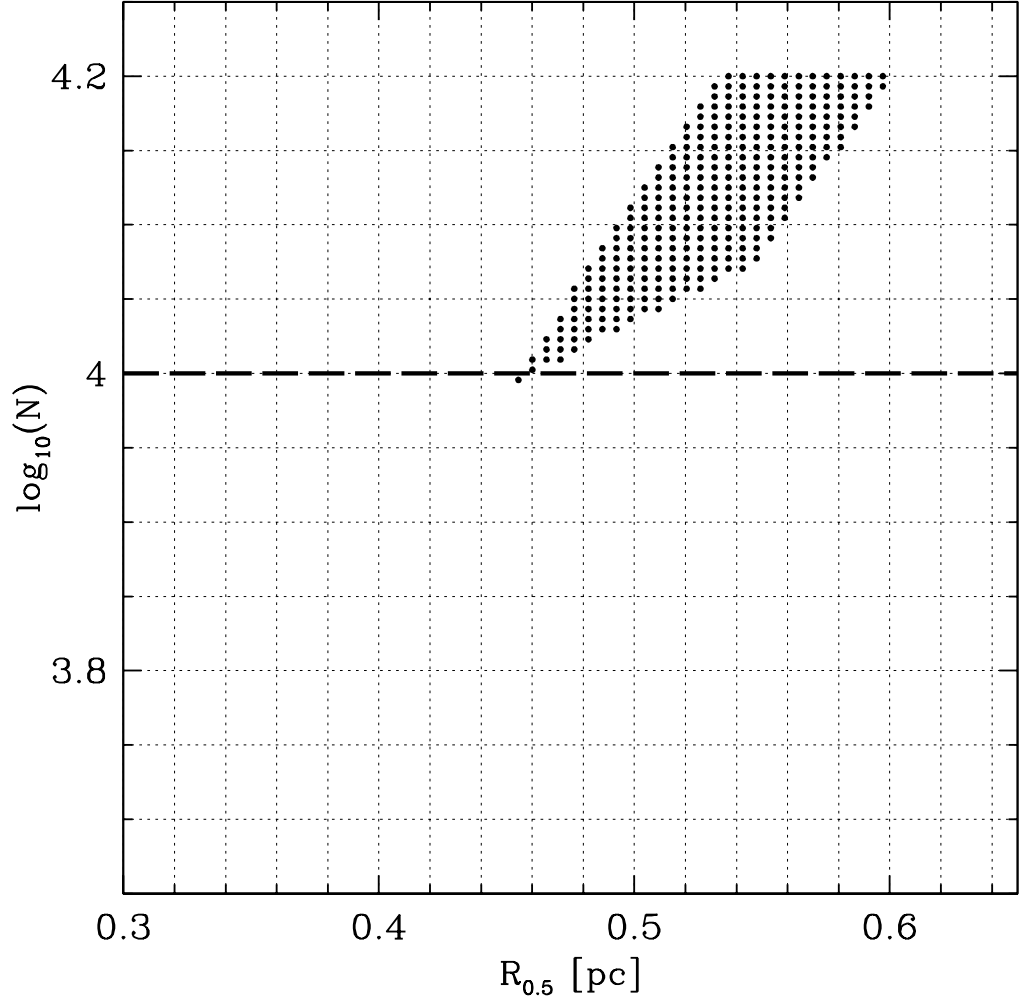}{17cm}{0}{70}{70}{-225}{-15}
\caption{The full set of candidate initial cluster models in virial
equilibrium.  The solution set is obtained by constructing a grid of
models with different $R_{0.5}$ and $N$, and selecting only those that
simultaneously satisfy $2.27\le \sigma_{\rm 1D} \le 2.81$~km/s,
$23.6\le n_{\rm c} \le 34.4$~stars and $t_{\rm cross}\le0.50$~Myr,
where $n_{\rm c}$ is the number of stars in the central volume with
radius $R=0.053$~pc. The boundaries on $\sigma_{\rm 1D}$ are the
one-sigma uncertainty range on the observed value. The limits on
$n_{\rm c}$ are $29\pm \sqrt{29}$.  Note that $N\ge10^4$ must also be
excluded because this many stars were certainly not present in the ONC
when it was born.
\label{fig:scn_vir}}
\end{figure}

Relaxing the constant-central density constraint, the full set of
candidate solutions is obtained (Fig.~\ref{fig:scn_vir}). As already
apparent in Fig.~\ref{fig:sc_vir}, the set of presumed solutions is
{\it very narrow}, with $N\approx10^4, M_{\rm cl}\approx
4.4\times10^3\,M_\odot$ and $R_{0.5}\approx0.45$~pc. It has been known
for a long time that the velocity dispersion in the Trapezium Cluster
is super-virial, if only the number of detected stars is used to
estimate the cluster mass (e.g. Hillenbrand \& Hartmann
1998). However, the candidate solutions found here demonstrate that,
by taking into account the binary companions, the natural loss of
stars from the cluster through its dynamical evolution, and allowing
for a small fraction of stars not detected yet, a model of the initial
ONC is found that is remarkably close to virial equilibrium.

\subsection{Expanding clusters}
\label{sec:gen_exp}
\noindent
Model~B2 in KPM shows that if the Trapezium Cluster is expanding then
it is being observed now at a time when the velocity dispersion and
central density have not decayed so much as to make the system appear
like an association. Expansion must have begun a few tens of thousands
of years ago, and the initial $f_{\rm tot}$ must have been similar to
that in the Galactic field, and thus significantly below that in
Taurus--Auriga.  Model~B2 assumes a high star-formation efficiency of
$\epsilon=0.42$, which nevertheless leads to an essentially rapid free
expansion of the entire system (Kroupa \& Frink 1999). Indeed, Lada,
Margulis \& Dearborn (1984) demonstrate that, if $\epsilon<0.5$ and
gas is removed instantaneously, then an unbound, expanding association
results. The presence of a few O~stars in the Trapezium Cluster
suggests that the removal of gas occurred on a time-scale comparable
to or shorter than a crossing time.

The set of candidate solutions to an expanding Trapezium Cluster,
assuming instantaneous gas removal and virial equilibrium of the
stellar and gaseous system before gas expulsion, can be constrained by
noting that the pre-expansion central density and velocity dispersion
must have been larger than the presently observed values, $\rho_{\rm
c}(t=0)>\rho_{\rm c}^{\rm (obs)}$ and $\sigma_{\rm
1D}(t=0)>\sigma^{\rm (obs)}$, where $t=0$ refers to the gas-expulsion
time. The central density cannot be made arbitrarily large, since
binary systems with semi-major axes as large as $a=440$~AU exist in
abundance (section~2.3 in KPM). Thus, if each binary system needs a
volume with a radius $\gamma\,a$ to form, and the forming systems
touch, then the maximum number density would be
\begin{equation}
\rho_{\rm c}^{\rm max}={3\times2\over 4\,\pi\,(\gamma\,a)^3}, 
\label{eqn:maxdens}
\end{equation}
so that $\rho_{\rm c}^{\rm max}=10^{7.69} -
10^{4.69}$~stars~pc$^{-3}$ for $\gamma=1-10$.

In a freely expanding flow that is unhindered by self-gravity, the
velocity dispersion measured within a projected volume with a constant
radius $2\,R$ decreases with time according to
\begin{equation}
\sigma_{\rm 1D}(t) = {R \over t + R/\sigma_{\rm 1D}(0)},
\end{equation}
where $\sigma_{\rm 1D}(0)=\sigma_{\rm 1D}/\sqrt{\epsilon}$ is the
velocity dispersion before gas expulsion, and $\sigma_{\rm 1D}$ is, as
in Section~\ref{sec:gen_vireq}, the velocity dispersion of the stellar
cluster if it were in virial equilibrium and without gas
(eqn.~\ref{eqn:1dvd}). This equation ensures that the initial velocity
dispersion, $\sigma_{\rm 1D}(0)$, is arrived at, and essentially
measures the typical velocity, $R/t$, of the stars remaining in the
measurement area. The term $R/\sigma_{\rm 1D}(0)$ is the time-lag
until stars with typical velocity $\sigma_{\rm 1D}(0)$ leave the area.
The velocity-dispersion model-data corrected for the radial bulk flow
plotted in the bottom panel of fig.~6 in KPM fit this relation very
well with $R=0.21$~pc.  That the flow in the KPM models~B is free on
scales larger than a few tenths of a~pc is also demonstrated in Kroupa
\& Frink (1999).  The time at which $\sigma_{\rm 1D}(t_{\rm
exp})=\sigma^{\rm (obs)}=2.54\pm0.27$~km/s is
\begin{equation}
t_{\rm exp} = R\,({1\over \sigma^{\rm (obs)}} - {1\over 
              \sigma_{\rm 1D}(0)}).
\end{equation}

In a free radial expansion flow, the central density in a Plummer
density distribution should evolve as
\begin{equation}
\rho_{\rm c}(t) = {3\,N \over 4\,\pi\,R_{\rm pl}(t)^3}, 
\end{equation}
because the Plummer radius of the system increases with time as
$R_{\rm pl}(t) = 0.77\,R_{0.5} + \sigma(0)\,t$. However, the model
data plotted in the upper panel of fig.~6 in KPM show that the real
central density, evaluated within a radius $R=0.053$~pc, decreases
much less rapidly. This is due to stellar-dynamical interactions in
the binary-rich system leading to a part of the population decoupling
from the flow and forming a small bound cluster, as discussed in
section~5.2.1 in KPM. Cooling of the flow through some binary
destruction appears to be instrumental in forming this decoupled core,
as can be inferred by comparing the central densities of models~B1
and~B2 at $t>0.2$~Myr in fig.~6 of KPM. This mode of cluster formation
will be addressed in more detail in a future contribution. For the
present, it has to be accepted that the evolution of $\rho_{\rm c}(t)$
cannot be expressed analytically.

\begin{figure}
\plotfiddle{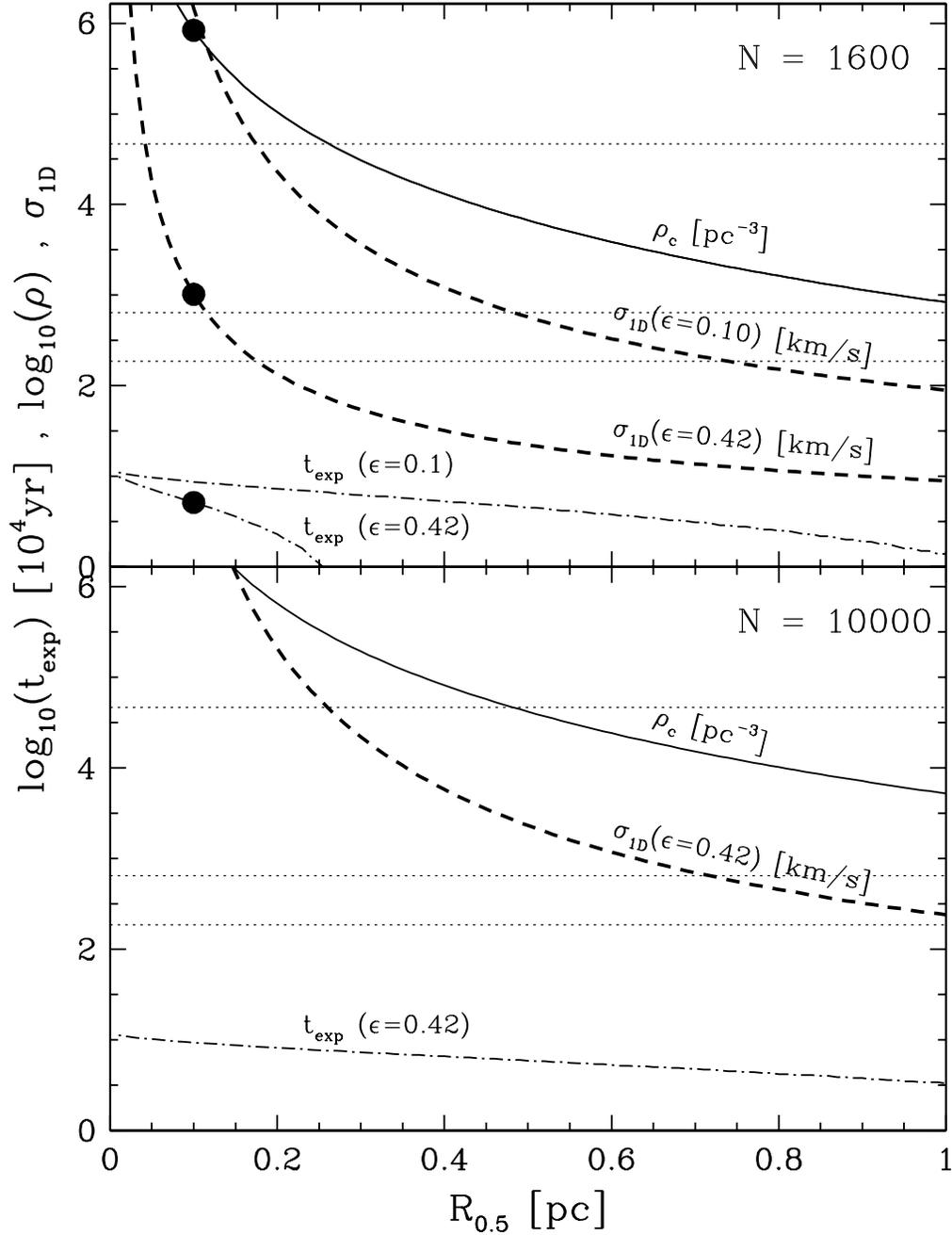}{17cm}{0}{70}{70}{-225}{-15}
\caption{Three sets of possible initial model clusters that expand as
a result of instantaneous gas loss. The upper panel displays two sets
of models for $N=1600$, and the lower panel shows one set of models
for $N=10^4$. For each model the initial central density,
log$_{10}(\rho_{\rm c})$, is plotted as the solid line, the
pre-expansion velocity dispersion, $\sigma_{\rm 1D}(0)$, is plotted
as thick short-dashed curves for star-formation efficiencies
$\epsilon=0.42$ and~0.10, and the time, log$_{10}(t_{\rm exp})$, when
$\sigma_{\rm 1D}(t_{\rm exp})=1.73$~km/s (the three-sigma lower
limit on the observed velocity dispersion) is shown as the dash-dotted
line. The lower three-sigma value is used to provide an upper boundary
on $t_{\rm exp}$ for each $R_{0.5}$. The one-sigma uncertainty range
on the observed velocity dispersion is represented by the two lower
horizontal dotted lines, and the upper horizontal dotted line is the
observed central density, log$_{10}(\rho_{\rm c}^{\rm (obs)})$ in the
Trapezium Cluster. For example, for model~B2, log$_{10}(\rho_{\rm
c})=5.92$~[stars/pc$^3$], $\sigma_{\rm 1D}(0)=3.01$~km/s and
log$_{10}(t_{\rm exp})=0.72$~[$10^4$~yr], i.e. $t_{\rm
exp}=5.2\times10^4$~yr (large black dots). This is the solution found
by KPM.
\label{fig:sc_exp}}
\end{figure}

A set of possible initial models is summarised in
Fig.~\ref{fig:sc_exp}.  Surmised solutions with $N=1600$ and a
star-formation efficiency of~42~per~cent have $R_{0.5}<0.12$~pc,
ensuring that the pre-expansion velocity dispersion is larger than the
observed value by at least a one-sigma error margin. Such models lead
to the velocity dispersion being consistent with the observed value
within $t\simless 10^5$~yr. For $\epsilon=0.1$, the candidate
solutions have $R_{0.5}<0.27$~pc, ensuring the pre-expansion central
density is larger than the currently observed value. In this case the
pre-expansion velocity dispersion is very large, requiring a longer
time to pass until the velocity dispersion has decayed to a value
consistent with that observed. Nevertheless, $t_{\rm
exp}<10^5$~yr. Candidate solutions with $N=10^4$ have
$R_{0.5}<0.47$~pc and $t_{\rm exp}\simless10^5$~yr for $\epsilon\le
0.42$.

\begin{figure}
\plotfiddle{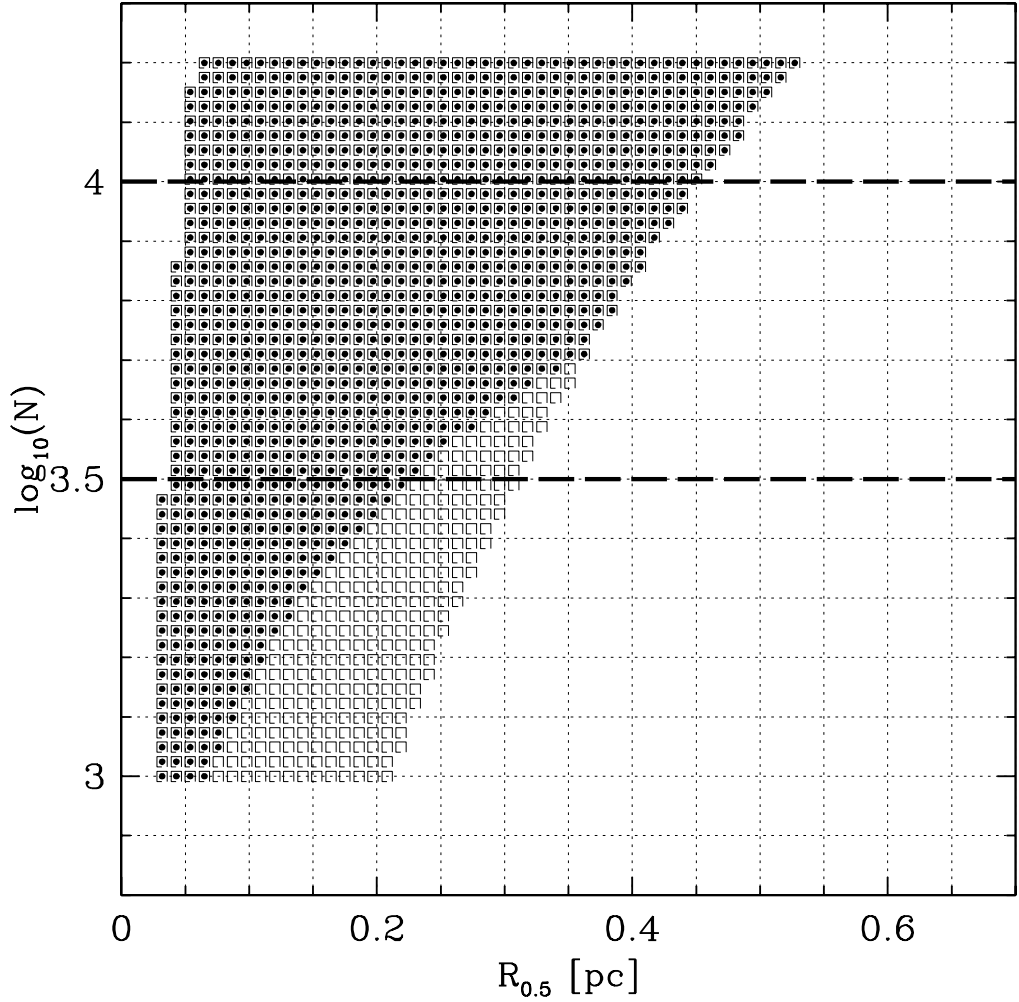}{17cm}{0}{70}{70}{-225}{-15}
\caption{The full set of candidate initial models for expanding
clusters. The candidate solutions satisfy simultaneously $\sigma_{\rm
1D}>2.81$~km/s, $n_{\rm c}>34.4$~stars (see Fig.~\ref{fig:scn_vir})
and $\rho_{\rm c}\le 10^{7.69}$~stars/pc$^3$.  Black dots are for a
star-formation efficiency $\epsilon=0.42$, and open squares are for
$\epsilon=0.1$.  Note that log$_{10}(N)<3.5$ should be excluded
because at least~3500 ONC stars have been seen, and $N\ge10^4$ must be
excluded because this many stars were certainly not present in the ONC
when it was born.
\label{fig:scn_exp}}
\end{figure}

The full set of candidate solutions is shown in
Fig.~\ref{fig:scn_exp}. For all candidate solutions, $t_{\rm
exp}<1.2\times10^5$~yr. The KPM calculations show that $f_{\rm tot}$
does not evolve significantly, so that $f_{\rm tot}\approx0.6$ is
required at the time when expansion ensues.

\subsection{Collapsing clusters}
\label{sec:gen_coll}
\noindent
An initially homogeneous sphere with radius $R_{\rm in}$ that starts
from rest collapses to a singularity under its own gravity within a
free-fall time, $t_{\rm col}\approx t_{\rm cross}/\sqrt{2}$
(e.g. Binney \& Tremaine 1987). If the initial matter distribution is
not smooth, for example when only a finite number of stars is present,
local density differences will lead to the growth of the tangential
velocity dispersion which limits the radius of the collapse to the
minimum value at the bounce (Aarseth, Lin \& Papaloizou 1988)
\begin{equation}
R_{\rm b}\approx R_{\rm in}/N^{1/3}.
\end{equation}
The post-collapse system exhibits a core-halo structure, and an
initial asphericity will be retained to some degree by the halo,
whereas the core will be approximately spherical (Boily, Clarke \&
Murray 1999). This may, in principle, be a model of the ONC, which has
an approximately spherical core (the Trapezium Cluster) and an
elliptical halo (Hillenbrand \& Hartmann 1998), provided the ONC is
old enough for the collapse to have progressed to the morphology
evident today.

The presence of a large number of primordial binary systems
complicates the situation dramatically because they have relatively
large interaction cross sections, and they constitute an energy sink
as well as a highly significant energy source. A cold collapse can be
envisioned being arrested when the densities become large enough that
the binary systems begin interacting. At this stage cooling through
binary disruption will reduce the relative velocities between the
systems, slowing down the collapse. Additionally, heating through
hardening binaries will also oppose the collapse. Vesperini \&
Chernoff (1996) address such issues by considering the collapse of an
initially homogeneous sphere with a binary proportion $f_{\rm
fot}=0.05$. They find that the primordial binaries do not alter the
collapse and violent relaxation. However, their results are not
applicable to the present work, since the binary proportion used here
is significantly higher, and because the results of Vesperini \&
Chernoff apply to the case where binary--binary collisions are
insignificant. Binary--binary collisions, however, dominate the
interactions in the realistic case considered here.  Details of the
processes await further research.

During a cold collapse starting from an initially spherical and
homogeneous configuration with initial (central) density $\rho_{\rm c,
in}$, the central density increases to a value at the bounce
$\rho_{\rm c, b} \approx N\times \rho_{\rm c, in}$, given the above
scaling for the linear dimension of the system.  This is not fulfilled
in models~C1 and~C2, since this scaling and eqn.~\ref{eqn:rhoc} for
$\rho_{\rm c, in}$, would imply $\rho_{\rm c, b}^{\rm C1} =
10^{7.3}$~[stars/pc$^3$] and $\rho_{\rm c, b}^{\rm C2} =
10^{6.5}$~[stars/pc$^3$], whereas from fig.~10 in KPM values
of~$10^{5.9}$ and~$10^{5.8}$~stars/pc$^3$, respectively, result in the
collapse computations. The central density at bounce is significantly
smaller because (i) the models have a small but finite initial
velocity dispersion, (ii) the initial density distribution is
centrally concentrated leading to a spread of arrival times near the
centre which increases $R_{\rm b}$, and (iii) binary systems oppose
the collapse.

Similarly, the one-dimensional velocity dispersion of the system at
bounce cannot be calculated from a simple scaling law, because at this
stage it is unknown how the binding energy of the binary-rich
initially centrally concentrated cluster evolves.  However, the data
in the lower panel of fig.~10 in KPM suggest that, if $\sigma_{\rm
vir, in}$ is the one-dimensional velocity dispersion a pre-collapse
cluster would have were it in virial equilibrium
(eqn.~\ref{eqn:1dvd}), then
\begin{equation}
\sigma_{\rm b} = 1.71 \times \sigma_{\rm vir, in}.
\label{eqn:vdb}
\end{equation} 
For models~C1 and~C2, the one-dimensional velocity dispersion
$\sigma_{\rm vir, in}=0.97$~km/s and~0.69~km/s (eqn.~\ref{eqn:1dvd}),
respectively, so that $\sigma_{\rm b}^{\rm C1} \approx 1.62$~km/s
and~$\sigma_{\rm b}^{\rm C1} \approx 1.20$~km/s.  Useful constraints
can be placed on the candidate solution-set of pre-collapse models of
the ONC if eqn.~\ref{eqn:vdb} is adopted for a comparison with
$\sigma^{\rm (obs)}$.

The one-dimensional velocity dispersion at bounce must be {\it larger}
than or equal to the observed value $\sigma_{\rm b}\simgreat
\sigma^{\rm (obs)}$. Also, the initial central density must be smaller
than or equal to the presently observed value, $\rho_{\rm c,in}\le
\rho_{\rm c}^{\rm (obs)}$. In addition, the collapse time must be
short enough to be consistent with the pre-main sequence age of the
Trapezium Cluster, $t_{\rm col}\simless 1$~Myr, since the cluster
could not have been collapsing stellar-dynamically for a time longer
than the age of the stars.  

The collapse time $t_{\rm col}\propto t_{\rm cross}$, where $t_{\rm
cross}$ is the crossing time of the pre-collapse cluster
(eqn.~\ref{eqn:tcr}). To estimate the constant of proportionality, the
data presented in fig.~10 of KPM need to be resorted to again. The
result is
\begin{equation}
t_{\rm col} = 0.97 \times t_{\rm cross}.
\end{equation}

\begin{figure}
\plotfiddle{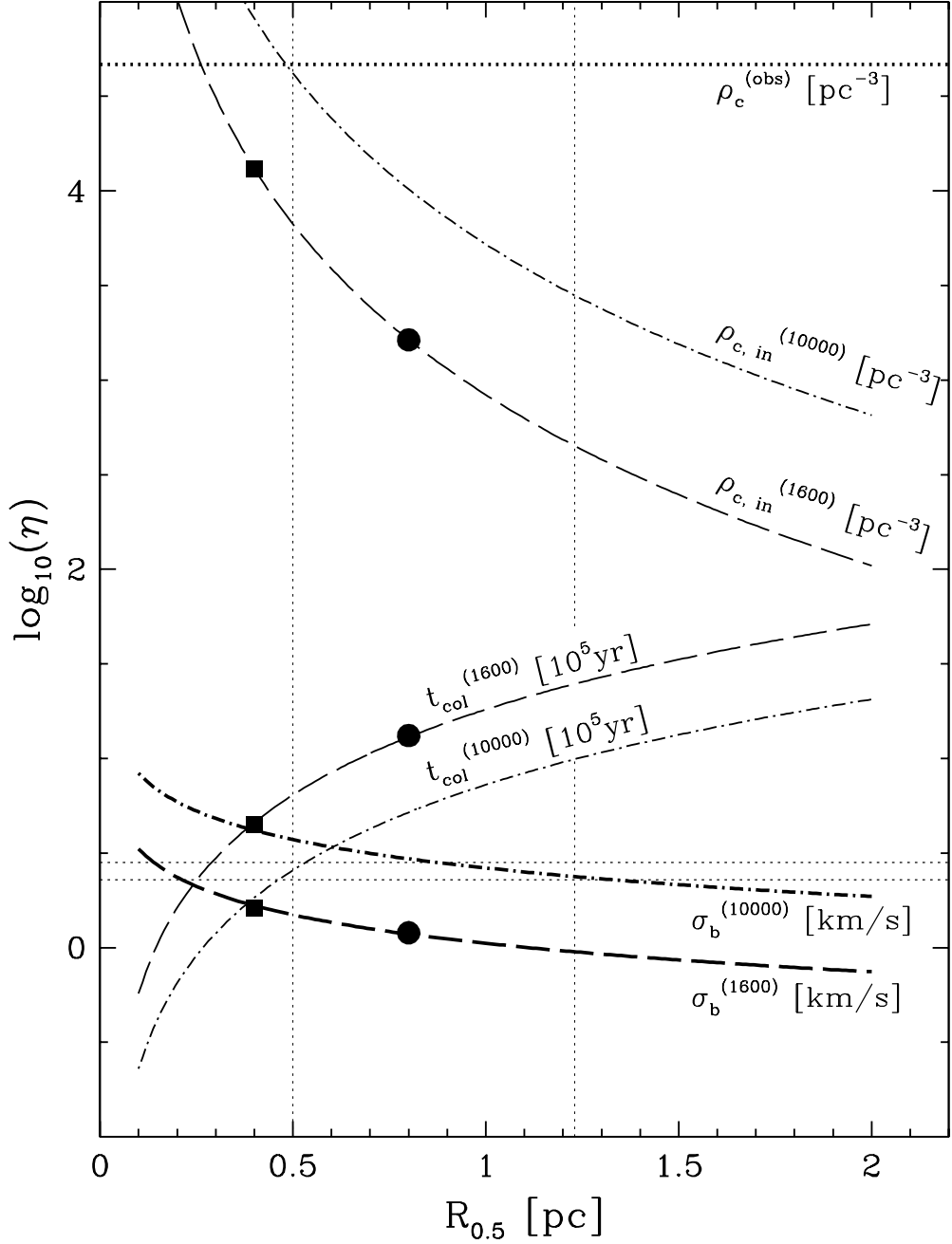}{17cm}{0}{70}{70}{-225}{-15}
\caption{Two sets of possible pre-collapse model clusters.
Long-dashed curves are for $N=1600$ and dash-dotted curves are for
$N=10^4$. For each, log$_{10}(\eta)$ is plotted in dependence of
$R_{0.5}$, with $\eta=\rho_{\rm c, in}, t_{\rm col}, \sigma_{\rm b}$.
For example, model~C1 has $\rho_{\rm c,
in}=1.3\times10^4$~stars/pc$^3$, and from fig.~10 in KPM, $t_{\rm
col}=0.45$~Myr and $\sigma_{\rm b}=1.62$~km/s (black squares), and
model~C2 has $\rho_{\rm c, in}=1.6\times10^3$~stars/pc$^3$, and from
fig.~10 in KPM $t_{\rm col}=1.32$~Myr and $\sigma_{\rm b}=1.20$~km/s
(black circles).  The upper horizontal dotted line is the presently
observed central density in the Trapezium Cluster, and the two lower
horizontal lines show the one-sigma range of the observed velocity
dispersion. The two vertical dotted lines delineate the candidate
solution set for $N=10^4$, the left boundary being determined by
$\rho_{\rm c, in}\le \rho_{\rm c}^{\rm (obs)}$ and the right boundary
by $\sigma_{\rm b} \ge \sigma^{\rm (obs)}$. Models with $0.5<
R_{0.5}<1.23$~pc are thus expected to satisfy the observational
constraints as they collapse with collapse times $t_{\rm
col}<10^6$~yr. There is no set of solutions for models with $N=1600$,
because when $\sigma_{\rm b} \ge \sigma^{\rm (obs)}$, $\rho_{\rm c,
in}$ is larger than the observed density.
\label{fig:sc_col}}
\end{figure}

Sequences of initial models are constructed for $N=1600$ and~$10^4$, and
$\rho_{\rm c, in}$, $t_{\rm col}$ and $\sigma_{\rm b}$ are calculated
for each as a function of $R_{0.5}$. The results are presented in
Fig.~\ref{fig:sc_col}. There is no candidate set of solutions for
$N=1600$, because the velocity dispersion at the bounce never reaches
the observed value for those cases when the central pre-collapse
density is smaller than the observed central density in the Trapezium
Cluster. However, there is a candidate solution set for models with
$N=10^4$. Such models have $0.5<R_{0.5}<1.2$~pc and should satisfy the
observational constraints at some time $\simless 1$~Myr during
collapse. Detailed $N$-body calculations are required to verify if this
is indeed a viable solution set, and if the central density and
velocity dispersion pass through the observational constraints at
about the same time.

\begin{figure}
\plotfiddle{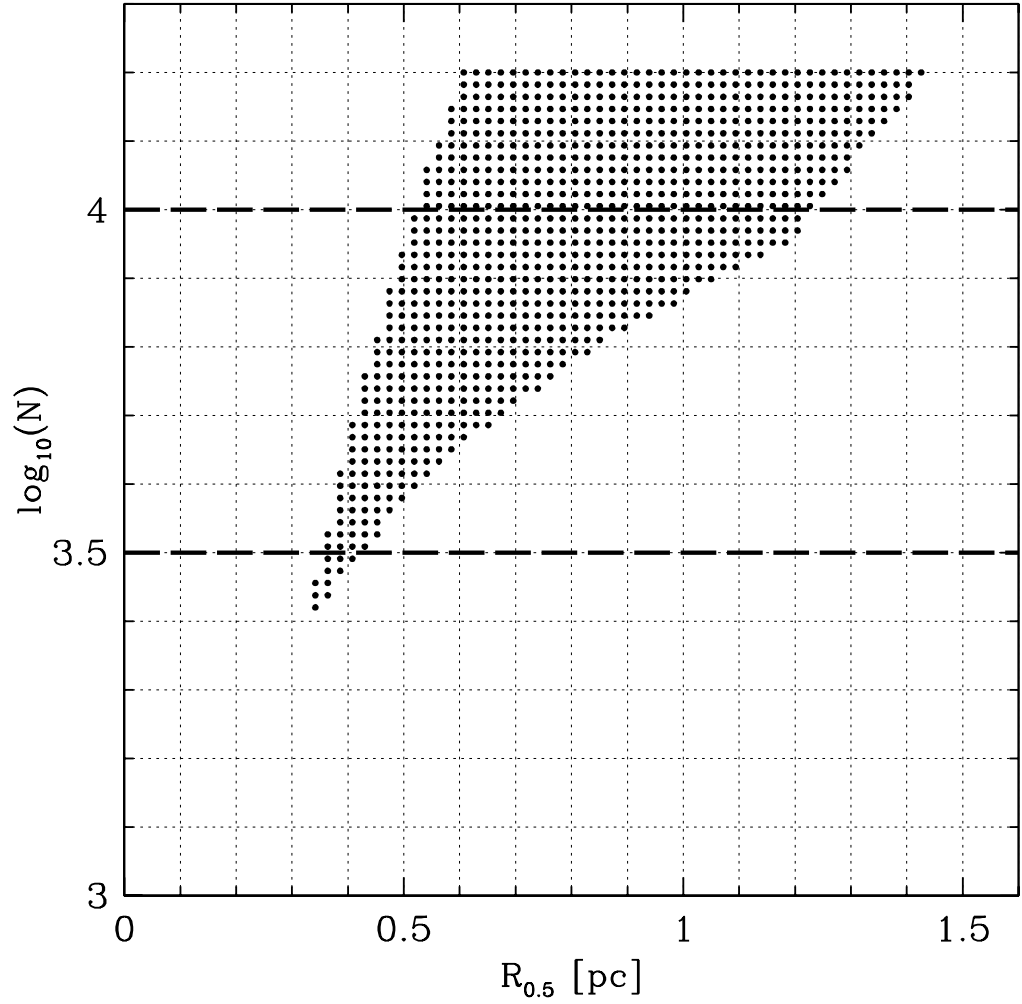}{17cm}{0}{70}{70}{-225}{-15}
\caption{The full set of candidate initial models of collapsing
clusters. The candidate solutions satisfy simultaneously $\sigma_{\rm
b}>2.27$~km/s, $n_{\rm c}<23.6$~stars (see Fig.~\ref{fig:scn_vir}) and
$t_{\rm col}<1$~Myr.  Note that log$_{10}(N)<3.5$ should be excluded
because at least~3500 ONC stars have been seen, and $N\ge10^4$ must be
excluded because this many stars were certainly not present in the ONC
when it was born.
\label{fig:scn_col}}
\end{figure}

The full set of candidate solutions is shown in
Fig.~\ref{fig:scn_col}. The results of KPM illustrate that $f_{\rm
tot}$ decreases significantly during $1-2\times t_{\rm col}$, but that
the binary proportion after virialisation depends on the initial
$R_{0.5}$. Thus, it is likely that initial $f_{\rm tot}<1$ (i.e. a
smaller binary proportion than in Taurus--Auriga) will be necessary
for presumed solutions with $R_{0.5}>0.5$~pc, to account for the
observed binary proportion in the Trapezium Cluster. Detailed
$N$--body calculations are needed to address this issue.

\subsection{Caveats}
\label{sec:cav}
\noindent
The models investigated here assume the clusters initially have
spherical and smooth Plummer density profiles. Deviations from this
will not change the results significantly for the same $\rho_{\rm c}$,
$M_{\rm cl}$ and $R_{0.5}$. A complication that will have to be
incorporated in future studies of the evolution of the ONC is the
tidal field exerted across it from the nearby molecular cloud. Kroupa
\& Frink (1999) show that the overall shape and velocity field of an
expanding young cluster can be affected noticeably in such a case.

\section{CONCLUSIONS}
\label{sec:conc}
\noindent
Using stellar-dynamical scaling laws together with results from
$N$-body calculations, three general classes of possible
stellar-dynamical solutions to initial configurations of the ONC (and
of the Trapezium Cluster) are investigated. By doing so, candidate
initial models have been found for all three classes. Thus, the ONC
may presently be (i) in or close to virial equilibrium, (ii)
expanding, or (iii) collapsing or in violent relaxation following a
cold collapse.  Class~(i) only allows a candidate solution in virial
equilibrium with $R_{0.5}\approx 0.45$~pc and $N\approx10^4$. Such a
large $N$ is realistic only if the binary proportion in the ONC
(rather than the Trapezium Cluster) is close to unity, and if about
30~per cent of stars are not detected and/or have already been ejected
to radii of a few~pc or larger.

The first and third classes will undoubtedly lead to a bound Galactic
cluster, while the second is likely to, on the basis of the formation
of a small bound cluster in the expanding Trapezium-Cluster models of
KPM.  The present results thus show that the ONC is most probably
forming a Galactic cluster. It is unclear though by which of the three
paths above this is occurring.

However, only {\it candidate} initial models have been presented
here. Follow-up, CPU-intensive $N$-body calculations will have to be
performed to verify which of these are consistent with the
observational constraints. The constraints include the velocity
dispersion, mass segregation, and the binary-proportion and density
profiles in the ONC.

\acknowledgements 
\vskip 10mm
\noindent{\bf Acknowledgements}
\vskip 3mm
\noindent
I am thankful to Chris Boily for useful correspondence, and Sverre
Aarseth for helpful discussions.  I acknowledge support from DFG
grant KR1635 and a short-visitor grant from the Smithsonian
Institution.

%

\clearpage

\clearpage
\newpage 

\clearpage
\newpage 

\clearpage
\newpage 

\clearpage
\newpage 

\clearpage
\newpage 

\clearpage
\newpage 

\end{document}